\documentclass[12pt,notoc]{JHEP3}

\usepackage{amsmath,amssymb,euscript,array,cite}

\input{epsf}
\usepackage{epsfig}

\newcommand {\slsh} [1] {\not{\hbox{\kern-2pt${#1}$}}}

\newcommand{\drawsquare}[2]{\hbox{%
\rule{#2pt}{#1pt}\hskip-#2pt
\rule{#1pt}{#2pt}\hskip-#1pt
\rule[#1pt]{#1pt}{#2pt}}\rule[#1pt]{#2pt}{#2pt}\hskip-#2pt
\rule{#2pt}{#1pt}}

\newcommand{\Yfund}{\raisebox{-.5pt}{\drawsquare{6.5}{0.4}}}
\newcommand{\Yasymm}{\raisebox{-3.5pt}{\drawsquare{6.5}{0.4}}\hskip-6.9pt%
                     \raisebox{3pt}{\drawsquare{6.5}{0.4}}%
                    }
\newcommand{\Ysymm}{\Yfund\hskip-0.4pt%
                    \Yfund}

\newcommand {\beq} {\begin{equation}}
\newcommand {\eeq} {\end{equation}}
\newcommand {\ber}{\begin{eqnarray*}}
\newcommand {\eer} {\end{eqnarray*}}
\newcommand {\bea}{\begin{eqnarray}}
\newcommand {\eea} {\end{eqnarray}}

\newcommand{\Dslash}{\,{\raise.15ex\hbox{/}\mkern-12mu D}}
\newcommand{\Cu}{\left( {\cal D} U_{\mu} \right)}
\newcommand{\Chiar}{\left(U_P\right)}

\title{Universality of $k$-string Tensions from Holography and the Lattice}
\author{Adi Armoni and Biagio Lucini \\
Department of Physics,\\ University of Wales Swansea,\\
Swansea, SA2 8PP, UK.\\
E-mail: {\tt a.armoni,b.lucini@swan.ac.uk}}
\preprint{SWAT/06/462}
\abstract{We consider large Wilson loops with quarks in higher representations in $SU(N)$ Yang-Mills theories. We consider representations with common ${\cal N}$-ality and check whether the expectation value of the Wilson loop depends on the specific representation or only on the ${\cal N}$-ality. In the framework of AdS/CFT we show that $\left\langle W_R \right\rangle  = \dim R \, \exp (-\sigma _k {\cal A})$, namely that the string tension depends only on the ${\cal N}$-ality $k$ but the pre-exponent factor is representation dependent. The lattice strong coupling expansion yields an identical result at infinite $N$, but shows a representation dependence of the string tension at finite $N$, a result which we interpret as an artifact. In order to confirm the representation independence of the string tension we re-analyse results of lattice simulations involving operators with common ${\cal N}$-ality in pure $SU(N)$ Yang-Mills theory. We find that the picture of the representation-independence of the string tension is confirmed by the spectrum of excited states in the stringy sector, while the lowest-lying states seem to depend on the representation. We argue that this unexpected result is due to the insufficient distance of the static sources for the asymptotic behaviour to be visible and give an estimate of the distance above which a truly representation-independent spectrum should be observed.}

\begin{document}

\section{Introduction}

In confining gauge field theories with fields in the adjoint representation, such as pure $SU(N)$ Yang-Mills theory or ${\cal N}=1$ Super Yang-Mills,
a tube of electric flux is expected to form between static charges. When $N>3$ an interesting problem is the evaluation of the expectation value of a Wilson loop
in a representation R,

 \beq 
\left\langle W_R \right\rangle \equiv \left\langle {\rm tr} \exp \left( i \int dx ^\mu A ^a _\mu T^a _R \right) \right\rangle . \label{problem}
\eeq

The common lore is that for large loops an area-law is expected and that the string tension $\sigma$ does not depend on the representation $R$ but only on its ${\cal N}$-ality $k$

\beq 
\left\langle W_R \right\rangle \sim \exp \left( -\sigma_k {\cal A} \right) \, .
\eeq

 The explanation is very simple and intuitive:
a cloud of gluons localised near the source will screen it, such that at a distance $\sim \Lambda _{\rm QCD} ^{-1}$ far from the source it will be impossible to tell what the original charge was and only its ${\cal N}$-ality $k$ can be associated with long distance observables.  If indeed the string tension is representation
 independent, we can choose the reducible representation $\Yfund \otimes \Yfund \otimes \, ... \, \otimes \Yfund $ (with $k$ boxes) to evaluate the $k$-string tension
 \beq 
\exp \left( - \sigma _k {\cal A} \right) \sim \left\langle {\rm tr} \exp \left( i \int dx ^\mu A ^a _\mu T^a _
 {\Yfund \otimes \Yfund \otimes \, ...\, \otimes \Yfund} \right) \right\rangle \, . 
\eeq
At infinite $N$, under the assumption of factorisation, we obtain
 \beq 
 \left\langle {\rm tr} \exp \left( i \int dx ^\mu A ^a _\mu T^a _
 {\Yfund \otimes \Yfund \otimes \, ... \, \otimes \Yfund} \right) \right\rangle =  \left\langle {\rm tr} \exp \left( i \int dx ^\mu A ^a _\mu T^a _
 {\Yfund} \right) \right\rangle ^k \sim \exp \left( -k \sigma _1 {\cal A} \right),
\eeq
namely $\sigma_k = k \sigma _1$. At finite $N$ we should think
about the $k$-string as a bound state of $k$ elementary strings. The dependence of the string tension on $k$ and $N$ is the subject of many papers \cite{Douglas:1995nw,Hanany:1997hr,Gross:1998gk,Herzog:2001fq,Hartnoll:2004yr,Lucini:2000qp,Lucini:2001nv,Lucini:2004my,DelDebbio:2001kz,DelDebbio:2001sj,DelDebbio:2003tk,Armoni:2003ji,Armoni:2003nz,Gliozzi:2005en,Gliozzi:2005dv,Shifman:2005eb} since it can teach us about the dynamics of the flux tube and reveal information on the mechanism of confinement.

The goal of this paper is to examine the issue of representation independence. We will analyse the problem from various angles: the AdS/CFT correspondence, the lattice strong coupling expansion and a lattice simulation. We will argue
that the string tension is indeed representation independent, but that the pre-exponent factor depends on the representation $R$. 

The organisation of the paper is as follows: in section 2 we analyse the problem using the Maldacena prescription \cite{Maldacena:1998im}. In section 3 we re-analyse the problem by using the lattice strong coupling expansion and we find at infinite $N$ a result which is identical to the holographic calculation. At finite $N$ we obtain an unexpected result: the string tension depends on the representation $R$ and not only on the ${\cal N}$-ality $k$. We interpret it as an artifact of the strong coupling expansion approach. Finally in section 4 we re-analyse the existing lattice data and we suggest that it supports universality: the string tension depends only on the ${\cal N}$-ality $k$.

\section{Wilson loops in the AdS/CFT correspondence}

In this section we wish to evaluate the expectation value of 
a Wilson loop with quarks in a higher representation.

Before we carry out the analysis let us state what is our expectation:
in a confining theory with dynamical fields in the adjoint representation
(such as ${\cal N}=1$ SYM) or with no dynamical fermions, such as pure Yang-Mills
theory, the string tension is expected to depend only on the ${\cal N}$-ality $k$ of the representation of the external quarks. It is, however, possible that the
pre-exponent factor will exhibit a dependence on the representation. Thus
the expectation is that in such theories, large Wilson loops will behave as

\beq 
\left\langle W_R \right\rangle \equiv \left\langle {\rm tr} \exp \left( i \int dx ^\mu A ^a _\mu T^a _R \right) \right\rangle =  F(R) \exp \left( -\sigma _k {\cal A} \right) \, .
\eeq
In particular in 2d YM it was found \cite{Gross:1993yt} that
\beq 
\left\langle W_R \right\rangle = \dim R \, \exp \left( -\sigma {\cal A} \right) \, ,
\eeq
namely that the pre-exponent factor is $\dim R$. The factor $\dim R$ is natural {\em at weak coupling}. In particular, when the gauge coupling is zero 
$ \left\langle W_R \right\rangle = \left\langle {\rm tr}\, 1 \right\rangle = \dim R$. 

In the AdS/CFT framework, the expectation value of a Wilson loop is the proper area, or the minimal surface of a string whose worldsheet boundary is the Wilson loop contour \cite{Maldacena:1998im}. The worldsheet boundary lies on the boundary of the AdS space where the field theory lives and extends to the interior of the AdS space.

We will consider Wilson loops in higher representation in a confining theory. In this case confinement manifests itself by the existence of an IR-cutoff in the geometry (either a black-hole horizon, or an 'end of space' situation). The string spends its time on the IR-cutoff and hence its proper area is proportional to the boundary area \cite{Brandhuber:1998er,Rey:1998bq}.

Our discussion is closely related to that of Gross and Ooguri \cite{Gross:1998gk} who considered
higher representations in ${\cal N}=4$. 

\subsection{String with $k=2$}

Consider the following three representations with $k=2$: the reducible representation $\Yfund \otimes \Yfund$, the antisymmetric representation $\Yasymm$ and the symmetric representation $\Ysymm$. Group elements transforming in those representations are related to the fundamental representation as follows
\bea
 & & \left\langle W_ {\Yfund \otimes \Yfund} \right\rangle = \left\langle ({\rm tr}\, U)^2 \right\rangle \nonumber \\
& &  \left\langle W_ {\Ysymm} \right\rangle = {1\over 2} \left ( \left\langle ({\rm tr}\, U)^2 \right\rangle +  \left\langle {\rm tr}\, U^2 \right\rangle \right ) \nonumber \\ 
& & \left\langle W_ {\Yasymm} \right\rangle = {1\over 2} \left ( \left\langle ({\rm tr}\, U)^2 \right\rangle -  \left\langle {\rm tr}\, U^2 \right\rangle \right ) \label{gt} \, , 
 \eea
where $U$ is a Wilson loop in the fundamental representation $U \equiv W_{\Yfund}$.

The above relations \eqref{gt} suggest a way to evaluate expectation values of Wilson loops in higher representation in terms of the expectation value of a Wilson loop in the fundamental representation. This is useful, since in string theory the fundamental string $F1$ is the only object that we have at hand\footnote{Recently, following Drukker and Fiol \cite{Drukker:2005kx}, several authors \cite{Hartnoll:2006hr,Yamaguchi:2006tq,Gomis:2006sb,Rodriguez-Gomez:2006zz} have used D3-branes and D5-branes to evaluate Wilson and Polyakov loops in higher representation of $SU(N)$ in ${\cal N}=4$ Super Yang-Mills.}.
 Higher representations will be obtained by products and sums of various configurations of the fundamental string. For example, the reducible two-index representation will be obtained by a product of two coincident closed strings, with a one winding each, corresponding to $\left\langle ({\rm tr}\, U)^2 \right\rangle$, see fig.~\ref{k2-1}.

\FIGURE[ht]{
\epsfig{width=4in,file=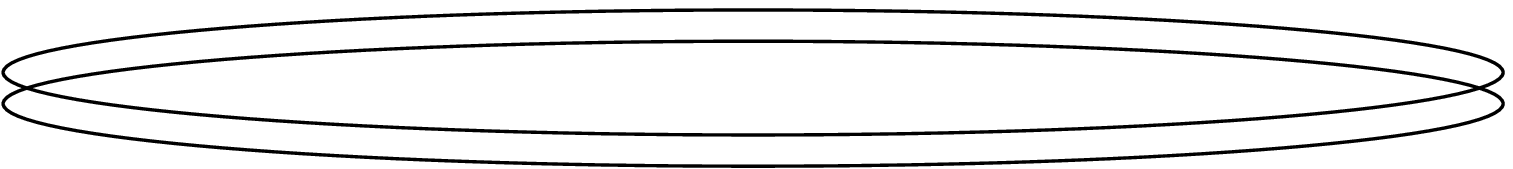}
\caption{$k=2$ string corresponding to the reducible representation.} \label{k2-1}
}

In the large $N$ limit we expect factorisation, namely
\beq
 \left\langle W_ {\Yfund \otimes \Yfund} \right\rangle = \left\langle ({\rm tr}\, U)^2 \right\rangle \rightarrow   \left\langle {\rm tr}\, U \right\rangle \left\langle {\rm tr}\, U \right\rangle \, .
\label{Wreducible}
\eeq 
Thus, the expectation value of the Wilson loop in the reducible representation is simply the square of the expectation value of the Wilson loop in the fundamental representation.

Next we discuss the case of the symmetric and the antisymmetric representations.
Due to \eqref{gt}, we propose that the expectation value of the symmetric 
(the antisymmetric) involves two contributions: a worldsheet with two coincident
fundamental loops plus (minus) a worldsheet of a string that winds twice, see fig.~\ref{k2-2}.

\FIGURE[ht]{
\epsfig{width=4in,file=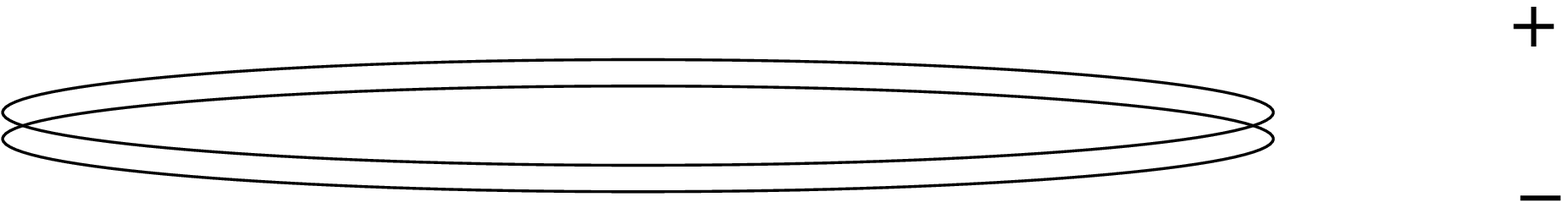}
\caption{$k=2$ strings: the symmetric and the antisymmetric representation.} \label{k2-2}
}

We argue that the expectation value of a Wilson loop in the fundamental
representation is 
\beq
\left\langle U \right\rangle = N \exp \left( -S_{\rm N.G.} \right) = N \exp \left( - \sigma _1 {\cal A} \right) \, .
\label{Wfund}
\eeq
Although in the original paper \cite{Maldacena:1998im} the pre-exponent factor $N$ was mentioned, it was not discussed in detail. In field theory it is expected, since the Wilson loop involves a trace over matrices in the fundamental representation. How can we understand it from the gravity side ?\footnote{We thank Carlos Nu\~nez for a discussion about this issue.} recall that the prescription of Maldacena \cite{Maldacena:1998im} for calculating Wilson loops involves Higgsing $U(N+1)\rightarrow U(N)\times U(1)$. In this way, a theory which contains dynamical matter in the adjoint representation can give rise to 'external' matter in the fundamental representation. In the AdS/CFT correspondence this is realized by separating a single brane from a stack of $N+1$ coincident branes. The separated brane can carry an arbitrary Chan-Paton factor and we have to sum over all possibilities, hence we get a factor $N$.

By using \eqref{Wreducible} and \eqref{Wfund} we find that
\beq
 \left\langle W_ {\Yfund \otimes \Yfund} \right\rangle =  N^2 \exp \left( -2S_{\rm N.G.} \right) = N^2 \exp \left( - 2\sigma _1 {\cal A} \right) \, ,
\eeq 
namely, that the string tension is $\sigma_2 = 2 \sigma _1$. This is expected since the $k=2$ string should be thought of a bound state of two fundamental string, but in the large $N$ limit (or $g_{st} \rightarrow 0$ limit) the two strings do not interact and hence its string tension is simply twice the tension of the fundamental string. 

Next we proceed to the case of symmetric/antisymmetric representations. Here we have to add/subtract the contribution from a string that winds twice. Consider a generic confining geometry of the form \cite{Kinar:1998vq}

\beq
 ds^2 = G_{uu} (du)^2 + G_{RR} (dR)^2 + G_{\Phi \Phi} (d\Phi)^2 + ...
\eeq
and a circular Wilson loop whose worldsheet coordinates are $r,\phi$. In the gauge $\phi = \Phi$, the Nambu-Goto action of a circular string that winds twice is simply twice the action of a string with with one winding
\bea
& & S_{\rm N.G.} ^{w=2} =  \nonumber \\
& &  {1\over 2\pi \alpha ' } \int _0 ^{4\pi} d\phi \int dr \sqrt {det\, G_{MN} \partial _ \alpha X^M \partial _\beta X^N} = \nonumber \\
& & {1\over 2\pi \alpha ' } \int _0 ^{4\pi} d\phi \int dr \sqrt {G_{\Phi \Phi}(r) (G_{RR}(r) + G_{uu}(r) (\partial _r u)^2) } = \nonumber \\
& &
 2 S_{\rm N.G.} ^{w=1}\, .
\eea

Accordingly,
\beq
 \left\langle W_ {\Ysymm} \right\rangle = {1\over 2} \left ( \left\langle ({\rm tr}\, U)^2 \right\rangle +  \left\langle {\rm tr}\, U^2 \right\rangle \right ) = {1\over 2} N^2 \exp \left( -2\sigma _1 {\cal A} \right) + {1\over 2} N \exp \left( -2\sigma _1 {\cal A} \right) \, ,
\eeq
and similarly,
\beq
 \left\langle W_ {\Yasymm} \right\rangle = {1\over 2} \left ( \left\langle ({\rm tr}\, U)^2 \right\rangle -  \left\langle {\rm tr}\, U^2 \right\rangle \right ) = {1\over 2} N^2 \exp \left( -2\sigma _1 {\cal A} \right) - {1\over 2} N \exp \left( -2\sigma _1 {\cal A} \right) \, .
\eeq
In all cases (with $k=2$) we find,
\beq
 \left\langle W_ R \right\rangle = \dim R \, \exp \left( -2\sigma _1 {\cal A} \right) \, ,
\eeq
namely an expected string tension $\sigma_2 = 2\sigma _1$ and a prefactor
that depends on the representation.

\subsection{An arbitrary representation}

We proceed now to the calculation of the expectation value of a Wilson loop
in an arbitrary representation $R$. 

We use the standard decomposition of a group element in representation $R$ in terms of the fundamental representation and use it for expressing the Wilson loop in a representation $R$ in terms of Wilson loops in the fundamental representation (see \cite{Gross:1998gk} for a discussion)
\beq
W_R \equiv {\rm tr} \exp \left( i \int dx ^\mu A ^a _\mu T^a _R \right) = \sum _ {P^n} C(P^n) \prod {\rm tr}\, U ^{w_i(P^n)} \label{dec}
\eeq
where the sum is over all possible partitions $P^n$ (all possible ways of writing an integer $\dim R$ in terms of a sum of positive integers). $C(P^n)$ are coefficients that depend on the partition $P^n$ and $w_i(P^n)$ are the elements of the partition. $w_i(P^n)$ must satisfy $\sum w_i(P^n) =k$.
From the limit of zero coupling we obtain $\dim R = \sum _ {P^n} C(P^n) N^{R(P^n)}$, where $R(P^n)$ is the number of elements (the number of traces in the product in \eqref{dec}) in the partition $P^n$.

The AdS/CFT predicts
\beq
\left\langle {\rm tr}\, U^w \right\rangle = N \exp \left( - w\sigma _1 {\cal A} \right)
\eeq
hence 
\beq
\left\langle W_R \right\rangle = \sum _ {P^n} C(P^n) N^{R(P^n)} \exp \left( - k \sigma _1 {\cal A} \right) = \dim R \, \exp \left( - k \sigma _1 {\cal A} \right) \, . \label{AdS-result}
\eeq
Thus the result \eqref{AdS-result} generalises the $k=2$ case. The AdS/CFT correspondence predicts the same tension for all representations with common ${\cal N}$-ality. The pre-exponent factor depends on the specific representation and the dependence is $\dim R$.

We expect that once the string coupling $g_{st}\equiv {1\over N}$ is turned on, namely when $1/N$ corrections are included, a result similar to \eqref{AdS-result}. In particular the pre-exponent factor looks like a kinematic factor. The dynamics
will modify the string tension, but only in powers of $1/N^2$, 
\beq
\sigma_k / \sigma_1 = k - {\cal O}(1/N^2) + ... \, , \label{N2}
\eeq
since the bulk dynamics is a closed string dynamics \cite{Armoni:2003ji,Armoni:2003nz}. So we conjecture the following expression at finite $k$ and $N$
\beq
\label{adscft}
\left\langle W_R \right\rangle = \dim R \, \exp \left( - \sigma _k  {\cal A} \right) \, , 
\eeq
namely that at finite $k$ and $N$ there is a prefactor which is exactly $\dim R$ (and in particular it does not depend on the Wilson loop contour) and an exponent with a string tension that satisfies \eqref{N2}.

\section{Wilson loops at strong coupling on the lattice}
The philosophy underlying Wilson loop calculations in the AdS/CFT framework shares similarities with the lattice strong coupling formalism, a well-established technique that exists for as long as the lattice approach itself\footnote{A good introduction to the lattice strong coupling formalism is given e.g. in~\cite{Smit:2002ug}.}. In this section, we provide a determination of the behaviour of the Wilson loop at leading order in the lattice strong coupling limit. Before proceeding, we note that the lattice strong coupling result is not expected to describe the continuum behaviour of Wilson loops, since the strong coupling series does not extrapolate beyond the bulk critical coupling~\cite{Lucini:2001ej,Lucini:2005vg}. However, a comparison of the lattice strong coupling and the AdS/CFT frameworks can give hints on features that could be universal properties of Wilson loops and hence manifest themselves in any formalism.

On the lattice, the vacuum expectation value ({\em vev}) of a Wilson loop in the representation $R$ of ${\cal N}$-ality $k$ is given by
\beq
\left\langle W_R \right\rangle = \frac{1}{Z} \int \Cu e^{- S} W_R \ ,
\eeq
where $\Cu$ is the gauge field measure defined over the link variables $U_{\mu}(i)$ and $Z = \int  \Cu e^{- S}$  
is the partition function of the theory. The choice of the lattice Yang-Mills action $S$ is not unique, since
lattice actions that differ by an $o(a^4)$ operator reproduce the same action in the limit $a \to 0$ ($a$ is
the lattice spacing). A natural choice for the problem at hand is the Wilson action
\beq
S = \beta \sum_P \left( 1 - \frac{1}{2N}  {\rm tr} (U_P + U_P^{\dag}) \right) \ , \label{Waction}
\eeq
where $\beta = 2N/g^2$ and $U_P$ is the parallel transport of links over the
elementary plaquette $P$ of the lattice.

Without loss of generality, we can assume that $R$ is an irreducible representation and the loop is planar. For simplicity, we also assume that $k \le N/2$. In order to compute the leading contribution to $\left\langle W_R \right\rangle$ at strong coupling we neglect the constant term in the action and perform an expansion of $e^{(\beta/2N)(U_P + U_P^{\dag})}$ in powers of $\beta/2N$. This allows us to rewrite the expression for the {\em vev} of the Wilson loop as 
\beq
\left\langle W_R \right\rangle = \frac{1}{Z} \int \Cu W_R \prod_P \sum_{n=0}^{\infty} \ \left(\frac{\beta}{2N}\right)^n \left( \sum_{m=0}^n \frac{1}
{m!(n-m)!} {\rm tr}\, U_P^m {\rm tr}\, U^{\dag n - m }_P \right) \ .
\label{wlexpansion}
\eeq
The terms that give a non-zero contributions in the expansion are those containing all links in the boundary appearing in powers such that their resulting ${\cal N}$-ality is opposite to the ${\cal N}$-ality of the Wilson loop. Thus, the minimal power of $\beta$ that gives a non-zero contribution in the sum over $n$ in~\eqref{wlexpansion} is $n = k$ and the terms that contribute is $\prod_{P \in {\cal A}}{\rm tr}\, U^{\dag k}_P$, where ${\cal A}$ is the area of the minimal surface bounded by the Wilson loop. Hence, for the purpose of computing the leading term in the strong coupling expansion, we can disregard the terms in the action containing ${\rm tr}\,U_P$.

Following the traditional route of strong coupling computations, we perform a character expansion of $e^{(\beta/2N) {\rm tr}\, U_P^{\dag}}$ 
\beq
\label{charexp}
e^{(\beta/2N){\rm tr}\, U_P^{\dag}} =  f + f \sum_r c_r d_r \chi^{\star}_r \Chiar \ ,
\eeq
where $\chi^{\star}_r \Chiar$ is the character (i.e. the trace) of $U_P^{\dag}$ in the representation $r$, $d_r$ is the dimension of that representation and the sum runs over all irreducible representations. Using the orthogonality relation of the characters 
\beq
\int \left( {\cal D} U \right) \chi_r^{\star}(U) \chi_{r^{\prime}}(U) = \delta_{r r^{\prime}} \ ,
\eeq
we get
\beq
d_r c_r = \frac{1}{f}\int \Cu e^{(\beta/2N) {\rm tr}\, U_P^{\dag}} \chi_r \Chiar\ ,
\eeq
with 
\beq
f  = \int \Cu e^{(\beta/2N) {\rm tr}\, U_P^{\dag}} \simeq 1 
\eeq
at the lowest order in $\beta$. By using the character expansion~\eqref{charexp}, it can be easily proved that in the strong coupling limit
\beq
\left\langle W_R \right\rangle = \dim R\, e^{- \sigma_R {\cal A}} \ , \qquad \sigma_R = - \log(c_R) \ ,
\eeq
with $\sigma_R$ expressed in lattice units.
This expression is similar to the AdS/CFT result~\eqref{adscft}. In particular, the factor in front of the exponential is the same, and is equal to the dimension of the representation. 

Let us now move on to the computation of $c_R$. The power expansions of
$e^{(\beta/2N) {\rm tr}\,U_P^{\dag}}$ in terms of the trace of the plaquette
in the fundamental representation $\chi_f \Chiar$ reads
\beq
\label{powerexp}
e^{(\beta/2N) {\rm tr}\,U_P^{\dag}} = \sum_{n=0}^{\infty} \frac{1}{n!}\left(\frac{\beta}{2N}\right)^n \chi_f^{\star n} \Chiar \ . 
\eeq
The ${\cal N}$-ality of $\chi_f^{\star n} \Chiar$ is $N - n$ mod(N). At the
lowest order in $\beta$ only the terms with $n \le N/2$ will matter. Since
$\chi_f^{\star n} \Chiar$ is the trace of an operator in a reducible
representation, it can be written as a weighted sum of traces of that operator
in irreducible representations of the same ${\cal N}$-ality:
\beq
\chi_f^{\star n} \Chiar = \sum_{r_n} m_{r_n} \chi^{\star}_{r_n} \Chiar \ ,
\eeq 
where $m_{r_n}$ counts how many times the irreducible representation $r_n$ of ${\cal N}$-ality $n$ enters the decomposition of the reducible representation given by the tensor product of $n$ ($n \le N/2$) fundamental representations. Inserting the previous equations back into~(\ref{powerexp}) and comparing the latter with~(\ref{charexp}), at the lowest order in $\beta$ we get
\beq
c_{r_k} = \frac{1}{k!}\left(\frac{\beta}{2N}\right)^k \frac{m_{r_k}}{d_{r_k}} \ .
\eeq
Since in the large $N$ limit $((k!) d_{r_k})/m_{r_k} = N^k$~\cite{Klink:1996ag}, in this limit we get 
\beq
c_{R} = \left(\frac{\beta}{2N^2}\right)^k \ ,
\eeq
which depends only on the ${\cal N}$-ality of $R$. Hence
\beq
\sigma_R = \sigma_k = - k \log \left(\frac{\beta}{2N^2}\right) = k \sigma_1 \ .
\eeq
The dependence of the string tension only on the ${\cal N}$-ality of the representation is in complete agreement with the AdS/CFT result. In the lattice strong coupling formalism, we also get an expression for $\sigma_k$ in terms of the (inverse) lattice 't Hooft coupling $\beta/(2N^2)$. It is not difficult to push forward the calculation and provide the first corrections to this expression at finite $N$, which comes from the exact value of $((k!) d_{r,k})/m_{r,k}$. For instance, for the irreducible representations of ${\cal N}$-ality two the leading $1/N$ correction to the string tension is $1/N$ for the symmetric and $-1/N$ for the antisymmetric representation. As this example shows, generally the leading correction will be in $1/N$ and not $1/N^2$, although exceptions to this rule might arise, as it is the case for the mixed symmetry irreducible representation with $k=3$. This result, at finite $N$, is surprising and unexpected: we do not expect $1/N$ corrections in pure $SU(N)$ Yang-Mills theory, since the 't Hooft genus expansion is in even powers of $1/N$ \cite{Armoni:2003ji,Armoni:2003nz}. Moreover, we expect the same string tension for the symmetric and the antisymmetric representation. We therefore interpret this as an artifact. In order to check the issue of universality we will re-analyse some lattice data in the next section. 

\section{Representation independence of $k$-strings and the lattice}
\label{lattice}
The independence of the string tension (or more in general of the string spectrum) from the specific representation at given ${\cal N}$-ality can be checked from first principles in SU($N$) Yang-Mills theory using lattice simulations. In lattice simulations, the most efficient way to extract string tensions is to use correlation functions of (multiply) winding strings. The correlators are expected to decay with the distance as the sum of exponentials:
\beq
\left\langle \Phi^{\dag}(0) \Phi(t) \right\rangle = \sum_i | c_i |^2 e^{- m_i x} \ ,
\eeq
where $m_i = \sigma_i L$ (L being the length of the loop and $\sigma_i$ the tension associated to the stringy state $|i\rangle$) and the coefficients $c_i = \left\langle 0 | \Phi^{\dag}(0) | 0 \right\rangle$, which give the overlap between $\Phi(0) | 0 \rangle$ and the eigenstate of the Hamiltonian $|i\rangle$, can be normalised in such a way that $\sum_i | c_i |^2 = 1$. Ideally a multiple exponential fit will provide information on the lowest-lying states of the spectrum, the highest ones dying on distances that are much smaller than one lattice spacing. It turns out however that this approach is inviable numerically. The most widely used technique to extract $\sigma_i$ is to build a set of $\Phi_{\alpha}$ and to find the linear combinations that make a given $|c_i|^2$ of order 1. In practice, a recursive procedure is used: from the correlation matrix $C_{\alpha \beta}(x) = \left\langle \Phi^{\dag}_{\alpha}(0) \Phi_{\beta}(x) \right\rangle$ the linear combination of $\Phi_{\alpha}$ which maximises the overlap with the ground state is worked out, then a complement space to this vector is built with the remaining trial operators and the procedure is repeated. This should allow to extract the mass of the next excitations. However, since the number of trial operators reduces as we go higher in the spectrum, the procedure becomes less and less efficient. Hence, while the method is very effective at determining the energy of the groundstate (provided that the trial operators have been judiciously chosen), it is less reliable in extracting the next excitations. Moreover, the higher the energy of a state the quicker the corresponding signal disappears as $x$ increases. Since statistical errors in correlation functions do not depend on $x$, this limits $x$ to 3 or 4 in practical cases. Hence, with the technique commonly used, typically one can confidently extract the energy of the first and second excited states, provided that their energy is no more than about two in units of the lattice spacing. (More details on the calculations of masses using correlation functions are provided e.g. in~\cite{Lucini:2004my}.)

\FIGURE[ht]{
\epsfig{scale=0.6,file=deg_su4_k2}
\caption{The masses of the lowest-lying states in the symmetric and antisymmetric representation of SU(4) on a $32^3$ lattice at $\beta = 60$.}
\label{figmc1}}
\FIGURE[ht]{
\epsfig{scale=0.6,file=deg_su6_k2}
\caption{The masses of the lowest-lying states in the symmetric and antisymmetric representation of SU(6) on a $24^3$ lattice at $\beta = 108$.}
\label{figmc2}}
\FIGURE[ht]{
\epsfig{scale=0.6,file=deg_su6_k3}
\caption{The masses of the lowest-lying states in the symmetric, mixed symmetry and antisymmetric representation of SU(6) on a $24^3$ lattice at $\beta = 108$.}
\label{figmc3}}

With this in mind, we look at the excited spectrum at fixed representation in SU(4) and SU(6) gauge theories in 2+1 and 3+1 dimensions. In the latter case, it has already been noticed in~\cite{Lucini:2004my} that states in representations other than the antisymmetric are all degenerate with states in the spectrum of the antisymmetric representation. For $k=3$ states in the symmetric spectrum are also degenerate with states in the mixed symmetry representation. This seems to indicate that the states in a given representation are all degenerate with states in all representations of lower symmetry.

The same pattern seems to emerge from the d=2+1 data. Using the Monte Carlo data discussed in Refs.~\cite{Lucini:2001nv,Lucini:2002wg}, we have analysed the spectrum in the k=2 symmetric and antisymmetric channels for SU(4) and SU(6) and in the k=3 symmetric, mixed symmetry and antisymmetric channels for SU(6) for the smallest lattice spacing simulated in both cases. The masses of the lowest-lying states of the spectrum identified in the analysis in each channel are plotted in figs.~\ref{figmc1},~\ref{figmc2}~and~\ref{figmc3} respectively. As in the d=3+1 case, the lowest state in the k=2 symmetric channel is degenerate with the first excited state in the k=2 antisymmetric channel for both SU(4) and SU(6) (as discussed in~\cite{Lucini:2004my}, by degeneracy here we mean that the splitting between the so-called degenerate states is much smaller than the typical splitting between states at fixed representation). In the k=3 sector we see a degeneracy between the lowest state in the mixed symmetry channel and the first excited state in the antisymmetric channel and the lowest state in the symmetric channel, the first excited state in the mixed symmetry channel and the second excited state in the antisymmetric channel. Things become less clear when we look at higher excitations, but this is to be expected, given that - as we have discussed before - the method for extracting energies of excited states is typically reliable only for the first few excitations.  

We interpret those observed features as a clear signature of the representation-independence of the spectrum: states extracted from operators in different representations that are degenerate in energy are in fact the same physical state, and not accidentally degenerate but different states. In fact, if the spectrum depends only on the ${\cal N}$-ality of the representation, we would expect to extract the same states, no matter which representation at given ${\cal N}$-ality we are considering. Hence the lattice spectrum reflects the universality of the string tension for representation with common ${\cal N}$-ality, as discussed in the previous sections.

A less intuitive and more surprising result is the observation that the lowest (stable) state in each representation does not appear on representations of higher symmetry. A possible explanation~\cite{DelDebbio:2003tk} could be as follows: the variational basis used for the symmetric representation has a small overlap with the true groundstate of the Hamiltonian and therefore correlation functions computed at large distance are needed in order to observe the decay of the symmetric quasistable string into the antisymmetric stable string. Although this seems unlikely at first, there are other examples in which a bad overlap hide a feature that is known to be present, like in the case of the absence of a peak in the specific heat for SU(2) gauge theory~\cite{Lucini:2005vg}.

The apparent absence of the lowest-lying states in the spectrum of representations with high symmetry can be explained with a simple model~\cite{Armoni:2003nz,Gliozzi:2005dv}. Consider for instance the correlation function of Polyakov loops in the symmetric two-index representation
\beq
\left\langle \Phi^{\dag}_{\Ysymm}(0) \Phi^{\dag}_{\Ysymm}(x) \right\rangle  = C \exp \left( -\sigma {\cal A} \right) + (1- C) \exp \left( -\sigma ^\star {\cal A} \right) \, .
\label{model}
\eeq 
where ${\cal A} = L \times x$ and $L$ is the length of the loop. By definition $\sigma ^\star > \sigma$, reflecting that $\sigma ^\star$ refers to the tension of the excited string state. It is clear that after a long time ($\Lambda _{\rm QCD} ^2 {\cal A}\gg 1$) the first term in \eqref{model} will dominate, namely the string will decay to its ground state. However, the actual area required for the decay is
\beq
(\sigma ^\star - \sigma) {\cal A} \gg \log { 1-C \over C} \,.
\label{size}
\eeq 
$C$ can be measured in lattice simulations by looking at the overlap of the state with tension $\sigma^{\star}$ with  $\Phi^{\dag}_{\Ysymm} | 0 \rangle$. A safe upper bound is $C \approx 0.2$. The two terms in~\eqref{model} have equal size when $(\sigma ^\star - \sigma) {\cal A} \approx  \log 4 \simeq 1.4 $. In current lattice calculations this bound is never reached. Among those accessible to us, the lattice calculations that uses the maximal area to date is the SU(4) d=2+1 set discussed in this paper. For this calculation, the effective string tension is most efficiently extracted by fitting correlators between 2 and 5 lattice spacings, hence ${\cal A} = (32 \times 3) a^2 = 96 a^2$. Since $\sigma ^\star - \sigma \simeq 0.01/a^2$, $(\sigma ^\star - \sigma) {\cal A} \approx 0.96$ and the bound~\eqref{size} is not fulfilled, despite our conservative assumptions. In order for the groundstate to be visible in this most favourable case fits to correlation functions should span at least over five lattice spacings. Due to the statistical noise, this is currently unfeasible with standard techniques.  

Our analysis can  be easily extended to the more general case of arbitrary $k$ and arbitrary representation. Unless our string state is the antisymmetric, an unbearably long simulation is needed to observe the groundstate. 

\section{Conclusions}

In this paper we analysed the issue of representation independence of large Wilson loops. By using holography and the lattice strong coupling expansion we found
that
\beq
\label{final-result}
\left\langle W_R \right\rangle = \dim R \, \exp \left( - \sigma _k  {\cal A} \right) \, . 
\eeq
A similar result was already derived in 2d Yang-Mills theory: the factor $\dim R$ appears in 2d as well, but in 2d the string tension depends on the representation, since gluons are non-dynamical in this model. Thus we can conclude that the factor $\dim R$ is universal and conjecture that it is an {\em exact} result in theories with adjoint fields.

In order to check whether the string tension $\sigma _k$ depends only on the ${\cal N}$-ality we also re-analysed some existing lattice data. Our analysis reasonably supports universality: representations with common ${\cal N}$-ality admit the same stringy spectrum. This conclusion is based on the assumption that operators with a given symmetry have a bad overlap with states with lower symmetry, so that in order to detect lowest-lying states using symmetric operators large distances must be reached. This is not in contradiction with the observed lattice spectrum. It would be interesting to repeat this analysis on the data used in Refs.\cite{Meyer:2004hv,KorthalsAltes:2005ph}, which exploit a new technique that allows to study correlation functions at larger distances.

Lattice simulations can also be used to check that the pre-exponent factor is indeed $\dim R$. Since the variational method used for extracting masses from correlation function is not designed for determining the prefactor, a dedicated calculation using different techniques would be needed.

\vspace{1cm}

{\bf Acknowledgements:}  The analysis presented in section~\ref{lattice} is based on data obtained by B.L. in collaboration with Mike Teper. We thank M. Creutz, S. Hands, H. Ita, T. Hollowood, C. Nu\~nez, M. Shifman and A. Rago for discussions. A.A. is supported by the PPARC advanced fellowship award. B.L. is supported by a Royal Society University Research fellowship.

\end{document}